\documentclass[a4paper,11pt]{article}
\usepackage{amsmath, amssymb, graphicx, physics, bm}
\usepackage[a4paper, margin=1in]{geometry}
\usepackage{tikz}
\usetikzlibrary{arrows.meta,calc}
\usepackage{jheppub, appendix, mathrsfs} 
\usepackage{lineno}

\title{\boldmath A Schwinger-Keldysh Formulation of Semiclassical Operator Dynamics}

\author{Jeff Murugan and Hendrik J. R. van Zyl}
\affiliation{The Laboratory for Quantum Gravity \& Strings,\\
Department of Mathematics and Applied Mathematics,\\
University of Cape Town, Private Bag, Rondebosch, 7701,\\
South Africa}

\emailAdd{jeff.murugan@uct.ac.za}
\emailAdd{hjrvanzyl@gmail.com}

\abstract{Krylov complexity has emerged as a useful diagnostic of operator growth in quantum many-body systems, complementing out-of-time-order correlators and spectral probes of chaos. Its standard formulation, based on Lanczos recursion and orthogonal polynomials, enables efficient computation but offers limited insight into the dynamical principles underlying operator spreading. In this work we develop a real-time Schwinger–Keldysh formulation of Krylov dynamics that treats Krylov complexity as an in–in observable generated by a closed-time-contour path integral. The resulting generating functional exposes an emergent phase-space description in which the Lanczos coefficients define an effective Hamiltonian governing operator motion along the Krylov chain. In the semiclassical limit, exponential complexity growth arises from hyperbolic trajectories, and asymptotically linear Lanczos growth appears as a universal chaotic fixed point, with subleading deformations classified as irrelevant, marginal, or relevant perturbations. Going beyond the saddle, the Schwinger–Keldysh framework provides controlled access to fluctuations and large deviations of Krylov complexity, revealing sharp signatures of integrability–chaos crossovers that are invisible at the level of the mean. This formulation reorganizes Krylov complexity into a dynamical field-theoretic framework and identifies new fluctuation diagnostics of operator growth in closed quantum systems.
}

\begin{document}
\maketitle
\flushbottom


\section{Introduction.}
\noindent
Understanding how quantum operators grow under time evolution is central to modern studies of quantum chaos, thermalization, and information scrambling \cite{Roberts:2018mnp, Chen:2018hjf, Parker:2018yvk, Bhattacharjee:2022vlt}. While early diagnostics of chaos focused on spectral statistics and level repulsion \cite{Bohigas:1983er, Ralf1999, Richter:2001ojn}, it has become increasingly clear that genuinely dynamical probes are required to capture the mechanisms by which information spreads through a many-body system. Among these, out-of-time-order correlators (OTOCs) \cite{Larkin1969QuasiclassicalMI, Maldacena:2015waa} have played a prominent role by quantifying the growth of operator commutators and establishing connections to classical chaos and Lyapunov behavior .\\

\noindent
Krylov complexity has emerged as a complementary and, in many respects, more structural diagnostic of operator growth \cite{Parker:2018yvk, Balasubramanian:2022tpr, Nandy:2024evd, Rabinovici:2025otw, Baiguera:2025dkc} . Rather than probing noncommutativity directly, Krylov complexity measures how a Heisenberg operator spreads in the Krylov basis generated by successive applications of the Liouvillian. Starting from a seed operator, one constructs an orthonormal basis by repeatedly commuting with the Hamiltonian and orthogonalizing, leading to a tridiagonal representation of the Liouvillian characterized by the Lanczos coefficients \cite{Lanczos1950AnIM, viswanath2008recursion}.\\ 

\noindent
The Krylov complexity, defined as the expectation value of the Krylov index under time evolution, provides a coarse but robust measure of how far the operator has propagated along this basis. One of the strengths of Krylov complexity is that it reorganizes operator growth into a one-dimensional dynamical problem. The entire many-body dynamics relevant for operator spreading is encoded in the Lanczos coefficients, which play the role of position-dependent hopping amplitudes along an emergent Krylov chain. This structure has led to a number of striking results, including bounds on operator growth \cite{Parker:2018yvk,Rabinovici:2020ryf, Avdoshkin:2022xuw}, connections between linear growth of Lanczos coefficients and exponential complexity growth, and deep links to spectral measures and random matrix theory.\\

\noindent
At present, essentially all quantitative studies of Krylov complexity rely on the Lanczos algorithm and its intimate connection to orthogonal polynomials and continued fractions \cite{Caputa:2021sib, Balasubramanian:2022tpr, Muck:2022xfc, Nandy:2024evd}. These methods are exceptionally efficient: they allow for high-precision numerical computation of Krylov complexity in large systems and yield exact analytic results in special cases where the spectral density is known. For many purposes, particularly in closed systems with well-controlled Hamiltonians, this approach is optimal. However, the very efficiency of the Lanczos formalism comes at a conceptual cost. By framing Krylov dynamics entirely in terms of recursion relations and spectral measures, the standard approach obscures the underlying dynamical structure governing operator propagation along the Krylov chain. In particular, it becomes difficult to address questions that are naturally phrased in dynamical or field-theoretic language. These include:
\begin{itemize}
    \item Universality across models:
    While empirical and analytic evidence suggests that broad classes of chaotic systems exhibit similar Krylov growth behavior (e.g., linear Lanczos coefficients leading to exponential complexity \cite{Parker:2018yvk}), the Lanczos formalism offers limited insight into why this universality arises or how it can be systematically classified.
	\item Finite-size effects and saturation:
    In realistic systems, exponential growth must eventually cross over to saturation due to finite Hilbert-space dimension. Understanding this crossover requires control over dispersion, boundary effects, and fluctuations—features that are difficult to isolate in the recursion-based approach.
	\item Coupling to environments:
    In open quantum systems \cite{Liu:2022god, Bhattacharya:2023zqt} the Liouvillian becomes non-Hermitian, the Lanczos recursion is no longer unique, and the notion of Krylov coefficients becomes basis-dependent. As a result, extending Krylov complexity beyond closed, unitary dynamics requires ad hoc modifications of the standard formalism.
\end{itemize}
\noindent
These limitations suggest that Krylov complexity, while computationally well-defined, lacks an explicit formulation as an effective dynamical theory. The central goal of this work is to provide such a formulation.\\

\noindent
In this article we show that Krylov dynamics admits a natural and universal description in terms of a real-time path integral. Starting from the tight-binding representation of the Liouvillian on the Krylov chain, we construct a Schwinger–Keldysh generating functional for Krylov observables and derive a phase-space path integral governing operator growth. In this formulation, the Lanczos coefficients determine an emergent effective Hamiltonian on Krylov phase space, revealing a semiclassical structure that is implicit, but hidden, in the standard recursion approach. The resulting framework reproduces familiar Krylov dynamics at the level of the mean complexity, while at the same time providing access to information well beyond that  captured by the mean alone.
A key advantage of the Schwinger–Keldysh formulation is that its generating functional encodes the full counting statistics of the Krylov position. It therefore determines not only the mean Krylov complexity, but its fluctuations and higher cumulants. This in turn reveals qualitative distinctions between systems whose mean complexity exhibits superficially similar growth.\\ 

\noindent
The path-integral formulation proposed in this article is not intended to replace the Lanczos algorithm as a computational tool. Instead, it reorganizes Krylov dynamics into an effective theory in which mean growth, fluctuations, universality, and deformations can be studied systematically. Exponential complexity growth, for example, emerges from hyperbolic classical trajectories when the Lanczos coefficients grow linearly, while deviations from linear growth translate into controlled corrections to the effective action. More generally, the generating functional provides a natural starting point for classifying operator growth not only by the asymptotic behavior of $K(t)$, but by the scaling of its entire cumulant hierarchy. The same framework also extends naturally to open systems through the inclusion of influence functionals on the closed time contour, providing a unified description of Krylov dynamics in closed and open quantum systems. By elevating Krylov complexity from a recursion-based construct to a dynamical effective theory, we aim to both clarify the geometric origin of operator growth and  expose physical information that is inaccessible when complexity is characterized by its mean alone.


\section{Krylov dynamics as a tight-binding problem.}
\label{Krylov-review}
\noindent
In this section we review how operator dynamics under Heisenberg evolution can be recast as an effective single-particle problem on the Krylov chain.   While much of this material is standard (see e.g. \cite{Parker:2018yvk, Caputa:2021sib}), we present it in a form that makes the subsequent path-integral formulation as transparent as possible.\\

\noindent
Let $H$ be a Hamiltonian acting on a Hilbert space $\mathscr{H}$, and let $O_0$ be a chosen seed operator. We consider the space of operators equipped with the Hilbert–Schmidt inner product
\begin{eqnarray}
    \langle A | B \rangle
    =
    \frac{1}{\mathcal N}\,\mathrm{Tr}(A^\dagger B),
\end{eqnarray}
where $\mathcal N$ is a normalization factor (typically the Hilbert-space dimension) chosen so that $\langle \mathbb I | \mathbb I \rangle = 1$. The generator of Heisenberg evolution is the Liouvillian superoperator $\mathcal L = [H,\cdot]$,
which is anti-Hermitian with respect to the Hilbert–Schmidt inner product for closed systems. Acting repeatedly with $\mathcal L$ on $O_0$ generates the Krylov subspace
\begin{eqnarray}
    \mathscr K = \mathrm{span}\{O_0,\,\mathcal L O_0,\,\mathcal L^2 O_0,\,\ldots\}\,.
\end{eqnarray}
Applying the Lanczos algorithm to $\mathcal L$ within this operator Hilbert space yields an orthonormal Krylov basis $\{|O_n\rangle\}$ with $|O_0\rangle \equiv O_0/\|O_0\|$, such that the action of the Liouvillian takes the tridiagonal form
\begin{eqnarray}
    \mathcal L |O_n\rangle
     =
     b_{n+1}|O_{n+1}\rangle
     +
     b_n|O_{n-1}\rangle\,,
\end{eqnarray}
with $b_0=0.$ For closed systems, the Lanczos coefficients $b_n$ are real and non-negative. They encode the entire structure of operator growth in the sense that once the $b_n$ are known, the evolution of any operator within the Krylov subspace is fully determined. Two points are worth emphasizing. First, the tridiagonal structure is not an approximation but a consequence of the orthonormalization. All higher-order couplings are absorbed into the definition of the basis. Second, while the Lanczos procedure depends on the choice of inner product, for closed systems the Hilbert–Schmidt inner product provides a natural and physically motivated choice, rendering the resulting $b_n$ uniquely defined.\\

\noindent
The Heisenberg-evolved operator $O(t) = e^{iHt}O_0 e^{-iHt}$ can be expanded in the Krylov basis as
\begin{eqnarray}
    |O(t)\rangle = \sum_{n\ge 0} \varphi_n(t)\,|O_n\rangle\,,
\end{eqnarray}
where the coefficients
$\varphi_n(t) = \langle O_n | O(t)\rangle$
may be thought of as a wavefunction on the non-negative integers $n=0,1,2,\ldots$.
Substituting this expansion into the Heisenberg equation $\partial_t |O(t)\rangle = -i\mathcal L |O(t)\rangle$
and using the tridiagonal action of $\mathcal L$, gives
\begin{eqnarray}
    i\,\partial_t \varphi_n(t)
    =
    b_{n+1}\,\varphi_{n+1}(t)
    +
    b_n\,\varphi_{n-1}(t)\,,
    \label{hopping}
\end{eqnarray}
with $\varphi_{-1}(t)=0$. This equation is mathematically identical to the Schrödinger equation for a single particle hopping on a semi-infinite one-dimensional lattice, with site index $n$ and nearest-neighbour hopping amplitudes $b_n$. The boundary at $n=0$ reflects the fact that the Krylov chain is generated from a single seed operator. Unlike conventional tight-binding models on an infinite lattice, the Krylov chain is only semi-infinite, with dynamics constrained by the $\varphi_{-1}(t)=0$ condition. There is no notion of ``negative complexity" and once the operator has spread away from $O_0$, it cannot propagate past the origin. In the tight-binding language, the boundary acts as a hard wall that enforces reflection of probability amplitude.\\

\noindent
The presence of this boundary has several important consequences. First, it ensures that the Krylov wavefunction remains normalizable and that $K(t) \ge 0$ at all times. Second, it plays a crucial role in finite-size and saturation effects. In finite-dimensional Hilbert spaces, the Krylov chain is effectively truncated at large $n$, so operator growth is confined between two boundaries. The resulting dynamics resembles wavepacket propagation in a finite box, with interference between reflections from the ultraviolet (large-$n$) cutoff and the infrared boundary at $n=0$. Finally, the boundary condition at $n=0$ singles out the seed operator as a distinguished point in Krylov space. In the semiclassical description developed later in this article, this boundary appears as an infrared regulator and is responsible for deviations from purely exponential growth at early times, before the wavepacket has moved sufficiently far from the origin for semiclassical approximations to become accurate. All in all, this tight-binding interpretation is a key conceptual launchpad. It shows that operator growth in a many-body Hilbert space can be recast as single-particle quantum mechanics on an emergent lattice, with all the complexity of the original system encoded in the position-dependent hopping coefficients $b_n$.\\

\noindent
Within this single-particle picture, we can define natural observables that characterize operator growth. The most basic of these is the position operator on the Krylov chain,
\begin{eqnarray}
    \hat n = \sum_{n\ge 0} n\,|n\rangle\langle n|\,.
\end{eqnarray}
The \textit{Krylov complexity} is defined as the expectation value of $\hat n$ in the evolving Krylov wavefunction,
\begin{eqnarray}
    K(t)
    =
    \langle \psi(t)|\hat n|\psi(t)\rangle
    =
    \sum_{n\ge 0} n\,|\varphi_n(t)|^2,
    \label{krylov-complexity}
\end{eqnarray}
where $|\psi(t)\rangle = \sum_n \varphi_n(t)|n\rangle$. Physically, $K(t)$ measures the typical Krylov index reached by the operator under time evolution and  provides a coarse but robust measure of operator growth. It is insensitive to microscopic details of the distribution $|\varphi_n(t)|^2$, making it well suited for diagnosing universal behaviour. More generally, the full probability distribution
\begin{eqnarray}
    P(n,t)=|\varphi_n(t)|^2\,,
\end{eqnarray}
defines the spread complexity. Higher moments of $P(n,t)$ encode fluctuations and dispersion of operator growth. For example, the variance
$\Delta n^2(t) = \langle \hat n^2\rangle - \langle \hat n\rangle^2$
measures the width of the Krylov wavepacket and distinguishes ballistic, diffusive, and localized propagation along the Krylov chain. Rare-event tails of $P(n,t)$ probe intermittency and large deviations, which are invisible at the level of the mean complexity. Most existing studies of Krylov complexity focus on $K(t)$ because it is readily accessible from Lanczos data and often displays simple growth laws. However, the recursion-based formalism provides no natural generating object for the full distribution $P(n,t)$. By contrast, once operator dynamics is recast as a genuine quantum evolution on the Krylov chain, it becomes natural to introduce generating functionals for moments and cumulants of $n(t)$.\\

\noindent
This distinction motivates the path-integral formulation developed in the following sections. By coupling a source to the Krylov position operator and working within a Schwinger–Keldysh framework, we gain systematic access not only to the mean complexity but also to fluctuations, large-deviation physics, and the effects of noise and dissipation on the entire distribution of operator growth.

\section{A brief review of the Schwinger-Keldysh formalism}
\label{sec:SK-review}
Before constructing a path integral for Krylov complexity, it is useful to briefly review the Schwinger–Keldysh (SK) formulation \cite{Keldysh:1964ud, Rammer:2007zz} of real-time quantum dynamics and to contrast it with the standard path-integral representation of transition amplitudes. Although the SK formalism is well established in nonequilibrium quantum field theory, its necessity and structure are often obscured in discussions focused on equilibrium or scattering observables.\\

\subsection{Managing expectation values}
The conventional real-time path integral computes \textit{transition amplitudes} between specified initial and final states. For a particle with Hamiltonian $H$, the fundamental object is
$K(x_f,t_f;x_i,t_i)=\langle x_f|e^{-iH(t_f-t_i)}|x_i\rangle$. By time slicing and inserting resolutions of identity, we obtain in a, by now, well known way
\begin{eqnarray}
    K(x_f,t_f;x_i,t_i)
=
\int_{x(t_i)=x_i}^{x(t_f)=x_f}\!\mathscr D x(t)\;
e^{\,iS[x]},
\end{eqnarray}
where $S[x]$ is the classical action for the particle dynamics.  This formulation is entirely adequate for computing amplitudes and scattering processes. However, it is intrinsically an in–out formalism in that it computes matrix elements between specified states at early and late times. It does not directly compute expectation values of operators at intermediate times.\\

\noindent
Many physical observables—transport coefficients, noise correlations, operator growth, and, as we argued above, Krylov complexity—are expectation values, not transition amplitudes. For example, the expectation value of the position operator for a particle prepared in an initial density matrix $\rho_0$, is given by
\begin{eqnarray}
    \langle \hat x(t)\rangle
    =
    \mathrm{Tr}\!\left(\rho_0\,e^{iHt}\hat x\,e^{-iHt}\right)\,.
\end{eqnarray}
This expression already reveals the essential structural difference from a transition amplitude; it involves both forward and backward time evolution. Writing it in the position basis makes this explicit,
\begin{eqnarray}
    \langle \hat x(t)\rangle
    =
    \int dx_f\,dx_i\,dx_i'\;
    \rho_0(x_i,x_i')\,
    \langle x_f|e^{-iHt}|x_i\rangle
    \langle x_i'|e^{+iHt}|x_f\rangle\,
    x_f.
    \label{SK-expectation}
\end{eqnarray}
In words, expectation values are built from a product of an amplitude and its complex conjugate, with $\rho_0$ gluing the two initial endpoints, and summed over the final state. No single-branch path integral can represent this structure. Instead, one must introduce two copies of the system, evolving forward and backward in time, and sew them together at the final time. This construction leads uniquely to the Schwinger–Keldysh or closed-time-path formulation.\\

\noindent
The central object of the Schwinger–Keldysh formalism is the generating functional
\begin{eqnarray}
    Z_{SK}[J_+,J_-]
    =
    \mathrm{Tr}\!\left(
    U_{J_+}(t,0)\,\rho_0\,U_{J_-}^\dagger(t,0)
    \right),
\end{eqnarray}
where
\begin{eqnarray}
    U_J(t,0) =
    \mathsf T
    \exp\!\left[
    -i\int_0^t dt'\,
    \big(H - J(t')\hat x\big)
    \right].
\end{eqnarray}
This object is the natural real-time generalization of the partition function. It is designed to generate expectation values of operators at finite time, rather than transition amplitudes between asymptotic states. The key structural feature is the doubling of time evolution;
$U_{J_+}(t,0)$ evolves the system forward in time from $0$ to $t$, while $U_{J_-}^\dagger(t,0)$ evolves it backward from $t$ to $0$. The trace over the final state ensures that the two evolutions are sewn together, producing an in–in observable rather than an in–out amplitude. The sources $J_+(t)$ and $J_-(t)$ couple to the operator $\hat x$ on the forward and backward branches of the contour, respectively. This doubling is not a technical redundancy but a logical necessity. A single source would not allow one to disambiguate operator insertions on the forward branch from insertions on the backward branch. But this distinction is essential in computing real-time expectation values, where forward and backward evolution play inequivalent roles. The two sources allow us to track insertions on each branch independently and to construct physically meaningful combinations. Moreover, setting $J_\pm=0$ immediately gives $Z[0,0] = \mathrm{Tr}(\rho_0) = 1$,
which guarantees automatic normalization. This is a major difference from the standard in–out path integral, where normalization must be imposed by hand and expectation values can suffer from spurious normalization factors.\\

\noindent
Expectation values are then obtained by functional differentiation. For example,
\begin{eqnarray}
    \langle \hat x(t)\rangle
    =
    \left.
    \frac{1}{i}
    \frac{\delta}{\delta J_q(t)}
    \ln Z[J_+,J_-]
    \right|_{J_\pm=0},
\end{eqnarray}
where $J_q = J_+ - J_-$. The appearance of the difference $J_q$ is crucial. Differentiation with respect to $J_+$ inserts $\hat x(t)$ on the forward branch, while differentiation with respect to $J_-$ inserts it on the backward branch with the opposite sign. Taking the difference ensures that the insertion corresponds to the physical operator evaluated at time $t$. By way of contrast, differentiation with respect to $J_c = (J_+ + J_-)/2$  generates response functions, and differentiation with respect to $J_+$ or $J_-$ alone does not correspond to a physical observable.\\

\subsection{Path integral representation}
\noindent
Given the ``amplitude $\times$ amplitude" structure of the Schwinger-Keldysh expectation value in \eqref{SK-expectation}, it is not surprising that the generating functional admits a path-integral representation in terms of the forward and backward trajectories,
\begin{eqnarray}
    Z_{SK}[J_+,J_-]
    =
    \int \mathscr D x_+\,\mathscr D x_-\;
    \exp\!\Big(
    iS[x_+] - iS[x_-]
    + i\!\int\!dt\,[J_+(t)x_+(t)-J_-(t)x_-(t)]
    \Big)\,.\nonumber\\
\end{eqnarray}
Here $x_+(t)$ and $x_-(t)$ represent the system’s coordinate on the forward and backward branches of the time contours. The appearance of the $+iS[x_+]$ and $-iS[x_-]$ reflects the forward and backward time evolution, respectively. The boundary conditions also follow directly from the operator definition of Z. In particular, the initial endpoints $x_+(0)$ and $x_-(0)$ are weighted by the matrix elements of the initial density matrix $\rho_0$. For a pure initial state this enforces $x_+(0)=x_-(0)$, while for a mixed state the two initial values are integrated independently with weight $\rho_0(x_+,x_-)$. The trace requires that the two paths coincide at the final time,
$x_+(t)=x_-(t)$, with an implicit integration over the common endpoint. This sewing condition ensures that the functional computes probabilities and expectation values rather than amplitudes.
Graphically, the contour runs forward from 0 to $t$ along the + branch and backward from $t$ to 0 along the - branch, forming a closed time loop. This closed contour is the defining feature of the Schwinger-Keldysh formalism and the reason it correctly computes real-time expectation values.\\

\noindent
It is often useful, especially in non-equilibrium physics, to rewrite the theory in terms of ``classical" and ``quantum" combinations,
\begin{eqnarray}
    x_c=\frac{x_+ + x_-}{2}\,,
    \quad
    x_q=x_+ - x_-\,.
\end{eqnarray}
Here $x_c(t)$ represents the average, physical trajectory, and $x_q(t)$ measures the difference between forward and backward paths and hence encodes quantum fluctuations. For a free particle, for example, with action
\begin{eqnarray}
    S[x]=\int_0^t dt\,\frac{m}{2}\dot x^2\,,
\end{eqnarray}
the Schwinger-Keldysh action
\begin{eqnarray}
    S[x_+] - S[x_-]
    =
    m\int_0^t dt\,\dot x_c\,\dot x_q
    =
    -\,m\int_0^t dt\,x_q\,\ddot x_c,
\end{eqnarray}
up to boundary terms that vanish under the sewing conditions. Integrating over $x_q$ imposes the condition $\ddot x_c = 0$,
which is precisely the classical equation of motion. Consequently, at the saddle-point level, the Schwinger-Keldysh path integral reproduces classical dynamics. Quantum fluctuations are encoded in deviations away from $x_q=0$. The same structure holds for the harmonic oscillator, where $\ddot x_c + \omega^2 x_c = 0$.\\

\noindent
This feature generalizes in an important way. In open systems \cite{Bhattacharyya:2026qef} or in the presence of noise, the Schwinger-Keldysh action acquires additional terms quadratic in $x_q$. Integrating out $x_q$ then yields stochastic equations of motion for $x_c$, with noise and dissipation arising from the same underlying influence functional. This is the precise sense in which this formalism provides a unified description of unitary dynamics, dissipation, and fluctuations.


\section{Schwinger--Keldysh formulation of Krylov complexity.}
\label{SK-Krylov}
As we have reviewed in Section \ref{Krylov-review}, once the Lanczos procedure has produced the tridiagonal representation, the dynamics of Krylov amplitudes is governed by the tight-binding Hamiltonian\footnote{We suppress possible diagonal terms for clarity; they may be included with no conceptual change.}
\begin{eqnarray}
    \hat H_{TB}
    =
    \sum_{n\ge 0} b_{n+1}\,\Big(|n\rangle\langle n+1|+|n+1\rangle\langle n|\Big)\,,
    \quad b_{n+1}\ge 0\,,
    \label{tight-hamiltonian}
\end{eqnarray}
acting on the Hilbert space spanned by the Krylov-chain basis states $\{|n\rangle\}_{n\ge 0}$. The initial condition corresponding to the seed operator is the state localized at the origin,
\begin{eqnarray}
    \rho_0 = |0\rangle\langle 0|,\qquad |\psi_0\rangle=|0\rangle\,.
\end{eqnarray}
The Krylov complexity is the expectation value of the position operator on the chain, $\hat n = \sum_{n\ge 0} n\,|n\rangle\langle n|$,
\begin{eqnarray}
    K(t)=\langle \psi(t)|\hat n|\psi(t)\rangle
    =\mathrm{Tr}\!\left(\rho_0\,\hat n(t)\right),
\end{eqnarray}
where the Heisenberg position operator $\hat n(t)=e^{i\hat H_{TB} t}\,\hat n\,e^{-i\hat H_{TB} t}.$ The key point here is that $K(t)$ is an in–in observable, and hence requires a closed time contour. To generate $\langle \hat n(t)\rangle$, we couple a source to $\hat n$ on each branch of the Schwinger-Keldysh contour. We define the Schwinger--Keldysh evolution operator generated by the Liouvillian,
\begin{eqnarray}
    U_J(t,0)
    =
    \mathsf T\exp\!\left[
    -i\int_0^t dt'\,\big(\mathcal L - J(t')\hat n\big)
    \right],
    \label{SK-Liouvillian-evolution}
\end{eqnarray}
where $\mathcal L=[H,\cdot]$ acts on the operator Hilbert space. The associated generating functional is
\begin{eqnarray}
    Z_{SK}[J_+,J_-]
    =
    \mathrm{Tr}\!\left(
    U_{J_+}(t,0)\,\rho_0\,U_{J_-}^\dagger(t,0)
    \right).
    \label{SK-Krylov-gen-fun}
\end{eqnarray}
\begin{figure}
\centering
\includegraphics[width=0.5\textwidth]{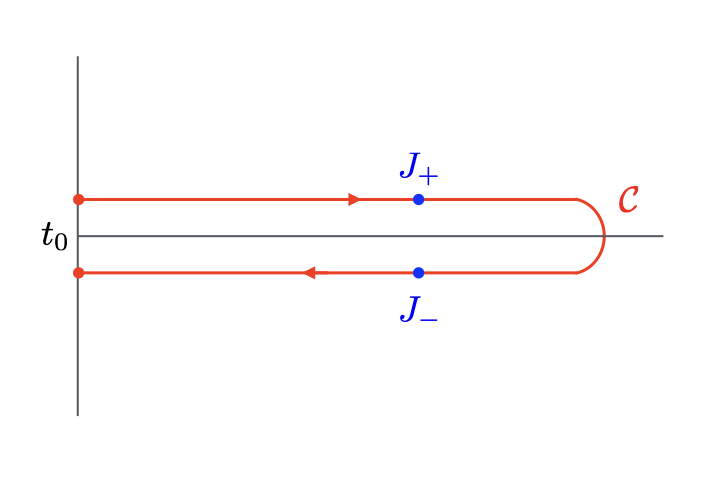}
\caption{Schwinger-Keldysh closed-time contour with insertions $J_+$ and $J_-$ on the forward and backward branches respectively.}
\label{SK-contour}
\end{figure}
This object has all the required properties. In particular, since $Z_{SK}[0,0]=\mathrm{Tr}(\rho_0)=1$, it has the correct normalization. Morover, the $J_+$ and $J_-$ independently generate insertions on the forward and backward branches. Finally, the physical observable is extracted by differentiating with respect to the quantum source
$J_q(t)=J_+(t)-J_-(t)$. In our case then,
\begin{eqnarray}
    \boxed{
    K(t)=
    \left.\frac{1}{i}\frac{\delta}{\delta J_q(t)}\ln Z_{SK}[J_+,J_-]\right|_{J_\pm=0}}
    \label{SK-Krylov-complexity}
\end{eqnarray}
At this level the formulation is completely basis independent and does not rely on the Lanczos construction. Once a Krylov basis $\{|n\rangle\}$ associated with the seed operator has been chosen, the Liouvillian admits a tridiagonal
representation,
\begin{eqnarray}
    \mathcal L \;\rightarrow\; \hat H_{TB}
    =
    \sum_{n\ge0} b_{n+1}(|n\rangle\langle n+1|+|n+1\rangle\langle n|),
\end{eqnarray}
and the Schwinger-Keldysh functional reduces to the tight--binding form above.\\ 

\subsection{Path integral representation}
\noindent
Discretizing time into $N$ slices each of size $\Delta t=t/N$, we write
\begin{eqnarray}
    U_{J_+}(t,0)\approx \prod_{k=0}^{N-1}
    \exp\!\left[-i\Delta t\,(\hat H_{TB} - J_{+,k}\hat n)\right]\,,  
\end{eqnarray}
where we have defined $J_{+,k}:= J_+(k\Delta t)$. A similar expression holds  for the backward branch. Next, insert complete sets of Krylov basis states 
\begin{eqnarray}
    \sum_{n_k\ge 0}|n_k\rangle\langle n_k|=\mathbb I\,,
\end{eqnarray}
between each short-time factor. For the forward branch, this yields a product of matrix elements $\langle n_{k+1}|
e^{-i\Delta t(\hat H-J_{+,k}\hat n)}
|n_k\rangle$. Since $\hat n$ is diagonal in the $|n\rangle$ basis, its contribution is immediate,
\begin{eqnarray}
    \langle n_{k+1}|
e^{-i\Delta t(\hat H-J_{+,k}\hat n)}
|n_k\rangle
=
e^{-i\Delta t(-J_{+,k}n_k)}\,
\langle n_{k+1}|e^{-i\Delta t\hat H}|n_k\rangle
\,+\,
\mathcal{O}(\Delta t^2)\,.
\end{eqnarray}
The central nontrivial ingredient then is the short-time hopping kernel
$\langle n_{k+1}|e^{-i\Delta t\hat H}|n_k\rangle$. To express this hopping kernel in a phase-space form, we introduce conjugate momentum states on the lattice. For a full integer lattice
\begin{eqnarray}
    |p\rangle=\sum_{n\in \mathbb Z}e^{ipn}|n\rangle\,,
\end{eqnarray}
with $p\in(-\pi,\pi]$. On the semi-infinite chain, this basis is not strictly orthonormal, but it remains a convenient semiclassical representation\footnote{In any case, we will only require the property that we can express a resolution of the identity in terms of these states.}. Equivalently we can continue $n$ to $\mathbb Z$ and enforce the boundary at $n=0$ by a method-of-images prescription. Either way, the result is a representation of the form
\begin{eqnarray}
    \langle n_{k+1}|e^{-i\Delta t\hat H}|n_k\rangle
    =
    \int_{-\pi}^{\pi}\frac{dp_k}{2\pi}\;
    \exp\!\left[ip_k(n_{k+1}-n_k)
    -i\Delta t\,H_{\mathrm{eff}}(n_{k+\frac12},p_k)\right]
    +\mathcal{O}(\Delta t^2)\,,\nonumber\\
\end{eqnarray}
where $H_{\mathrm{eff}}$ is the Weyl symbol of the hopping Hamiltonian and $n_{k+\frac12}=(n_{k+1}+n_k)/2$ is the midpoint prescription that makes the continuum limit well defined. For the hopping Hamiltonian above, the effective Hamiltonian
\begin{eqnarray}
    H_{\mathrm{eff}}(n,p)
    =
    b\!\left(n+\tfrac12\right)e^{ip}
    +
    b\!\left(n-\tfrac12\right)e^{-ip}.
\end{eqnarray}
Expanding for slowly varying $b(n)$ gives
\begin{eqnarray}
    H_{\mathrm{eff}}(n,p)
    =
    2b(n)\cos p
    +
    i b'(n)\sin p
    +
    \frac14 b''(n)\cos p
    +\cdots.
\end{eqnarray}
The leading term $2b(n)\cos p$ controls the semiclassical dynamics; the derivative corrections encode ordering and gradient effects. Finally, putting the forward and backward branches together, we obtain
\begin{eqnarray}
    Z_{SK}[J_+,J_-]
=
\int \mathscr{D}n_+\,\mathscr{D}p_+\,
\mathscr{D}n_-\,\mathscr{D}p_-\;
\exp\,\Bigl(
iS_{SK}[n_+,p_+;J_+]
-
iS_{SK}[n_-,p_-;J_-]
\Bigr)\,,\nonumber\\
\end{eqnarray}
with
\begin{eqnarray}
    S_{SK}[n,p;J]
    =
    \int_0^t dt'\,
    \Big(
    p\,\dot n
    -
    H_{\mathrm{eff}}(n,p)
    +
    J(t')\,n
    \Big)\,.
\end{eqnarray}
The boundary conditions on the integral are implemented by the initial state and the trace conditions. Specifically, since $\rho_0=|0\rangle\langle 0|$, it follows that $n_+(0)=n_-(0)=0$ while the trace enforces $n_+(t)=n_-(t)$, with an implicit sum over the common endpoint. Finally, to extract the Krylov complexity, we introduce classical/quantum fields
\begin{eqnarray}
    n_c=\frac{n_++n_-}{2},
    \quad
    n_q=n_+-n_-,
\end{eqnarray}
and similarly for $p_c,p_q$ in terms of which the generating functional reads
\begin{eqnarray}
    Z_{SK}[J_c,J_q]
    =
    \int \mathscr{D}n_c\mathscr{D}n_q
    \mathscr{D}p_c\mathscr{D}p_q\;
    \exp\!\left(iS_{\mathrm{K}}[n_c,n_q,p_c,p_q]+i\int dt\,J_q(t)\,n_c(t)+\cdots\right),\nonumber\\
\end{eqnarray}
where now $S_{K}$ is the Keldysh action obtained from the difference of branch actions. The precise form is not needed to extract $K(t)$; what matters is that
\begin{eqnarray}
    K(t)
    =
    \langle n(t)\rangle
    = \left.\frac{1}{i}\frac{\delta}{\delta J_q(t)}\ln Z[J_+,J_-]\right|_{J_\pm=0}
    =
    \langle n_c(t)\rangle.
\end{eqnarray}
At the saddle-point level ( at least for the unitary case), variation with respect to $n_q,p_q$ enforces classical Hamilton equations for $n_c,p_c$, with fluctuations and corrections obtained systematically by expanding around the saddle.

\section{Examples}
\label{examples}
Before proceeding any further, it is worth sanity-testing this Schwinger-Keldysh formulation on a range of systems whose Krylov complexities have already been computed using more conventional methods \cite{Caputa:2021sib, Balasubramanian:2022tpr, Muck:2022xfc, Chattopadhyay:2023fob}.

\subsection{Square-root hopping}
For our first example, we choose $b_{n+1}=g\sqrt{n+1}$. This choice of Lanczos coefficients is not as arbitrary as it seems. With it, the Krylov chain Hamiltonian is literally the harmonic oscillator position operator (up to a constant), so everything is analytic and tractable. Starting from the hopping Hamiltonian on the semi-infinite chain,
\begin{eqnarray}
    \hat H=\sum_{n\ge0} g\sqrt{n+1}\big(|n\rangle\langle n+1|+|n+1\rangle\langle n|\big)\,,
\end{eqnarray}
we introduce the harmonic oscillator ladder operators in the number basis $ \{|n\rangle\}$,
\begin{eqnarray}
    a|n\rangle=\sqrt{n}\,|n-1\rangle,\qquad a^\dagger|n\rangle=\sqrt{n+1}\,|n+1\rangle\,.
\end{eqnarray}
Then
\begin{eqnarray}
    a+a^\dagger=\sum_{n\ge0}\sqrt{n+1}
    \big(|n\rangle\langle n+1|+|n+1\rangle\langle n|\big),
\end{eqnarray}
so $\hat H = g\,(a+a^\dagger)$. Next, we take the seed state $|\psi_0\rangle=|0\rangle$. Then
\begin{eqnarray}
    |\psi(t)\rangle=e^{-i\hat H t}|0\rangle
    =e^{-igt(a+a^\dagger)}|0\rangle.
\end{eqnarray}
But if we define $D(\alpha):=\exp\left(-igt(a+a^\dagger)\right)=\exp\left(\alpha a^\dagger-\alpha^* a\right)$ with $\alpha=-igt$, then it is clear that
$|\psi(t)\rangle = |\alpha\rangle$ is just a usual harmonic oscillator coherent state. In other words, the Krylov amplitudes in the $|n\rangle$ basis are the standard coherent-state coefficients
\begin{eqnarray}
   \varphi_n(t)=\langle n|\alpha\rangle
   =e^{-|\alpha|^2/2}\frac{\alpha^n}{\sqrt{n!}}
   =e^{-g^2t^2/2}\frac{(-igt)^n}{\sqrt{n!}}\,. 
\end{eqnarray}
Hence the spread distribution is Poisson,
\begin{eqnarray}
    P(n,t)=|\varphi_n(t)|^2
=e^{-g^2t^2}\frac{(g^2t^2)^n}{n!}\,,
\end{eqnarray}
and the Krylov complexity is simply
\begin{eqnarray}
    K(t)=\sum_{n\ge0} nP(n,t)=g^2t^2\,.
\end{eqnarray}
This is a well-known quadratic growth law exhibited by many integrable systems. To recover this result in the Schwinger-Keldysh formulation we take $\rho_0=|0\rangle\langle0|)$ and 
\begin{eqnarray}
    Z_{SK}[J_+,J_-]=\mathrm{Tr}\big(U_{J_+}\rho_0 U_{J_-}^\dagger\big),
    \quad
    U_J=\mathsf T\exp\!\left[-i\int_0^t dt'\big(\hat H-J(t')\hat n\big)\right]\,,
\end{eqnarray}
with $\hat n=a^\dagger a$. The choice of source functions is a little subtle and so warrants a short digression. What we are trying to compute in this formulation is  not just the mean Krylov complexity
$K(t)=\langle \hat n(t)\rangle$,
but the entire probability distribution
$P(n,t)=\Pr(\hat n(t)=n)$,
or equivalently its generating function
\begin{eqnarray}
    Z(\chi)
    =
    \sum_n P(n,t)e^{i\chi n}
    =
    \big\langle e^{i\chi \hat n(t)}\big\rangle\,,
\end{eqnarray}
With the generating function in hand, we can easily compute
\begin{eqnarray}
    \langle \hat n(t)\rangle
    =
    \left.\frac{1}{i}\frac{\partial}{\partial\chi}\ln Z(\chi)\right|_{\chi=0},
    \quad
    \text{Var}(n)=
    -\left.\frac{\partial^2}{\partial\chi^2}\ln Z(\chi)\right|_{\chi=0},
\end{eqnarray}
and so on. Now, suppose we choose a source localized at the final time, 
$J(t')=\lambda\,\delta(t'-t)$. Since time ordering puts later operators to the left,
\begin{eqnarray}
    U_J(t,0)
    =
    e^{-i\hat H t}\,
    e^{+i\lambda \hat n}.
\end{eqnarray}
So a delta-function source at $t$ inserts a factor of $e^{+i\lambda \hat n(t)}$. To compute the generating function, we want one insertion of $e^{i\chi \hat n(t)}$ inside an expectation value,
\begin{eqnarray}
    \langle e^{i\chi \hat n(t)}\rangle
    =
    \mathrm{Tr}\!\left(
    \rho_0\,e^{i\hat H t}\,e^{i\chi\hat n}\,e^{-i\hat H t}
    \right).
\end{eqnarray}
However,the Schwinger-Keldysh functional has two evolutions, forward and backward. If we naively put the same source on both branches, we would double-count the insertion. Instead, we need to split the insertion symmetrically so that
\begin{eqnarray}
    J_+(t')=\frac{\chi}{2}\,\delta(t'-t),
    \quad
    J_-(t')=-\frac{\chi}{2}\,\delta(t'-t)\,.
\end{eqnarray}
On the forward branch, this results in 
\begin{eqnarray}
    U_{J_+}(t,0)
    =
    e^{-i\hat H t}\,e^{+i(\chi/2)\hat n}
\end{eqnarray}
 and 
\begin{eqnarray}
    U_{J_-}^\dagger(t,0)
    =
    \big(e^{-i\hat H t}e^{-i(\chi/2)\hat n}\big)^\dagger
    =
    e^{+i(\chi/2)\hat n}\,e^{+i\hat H t}\,,
\end{eqnarray} 
on the backward branch. Inserting both of these into the expression for the generating functional gives
\begin{eqnarray}
    Z(\chi)
    &=&
    \mathrm{Tr}\!\left(
    e^{-i\hat H t}\,e^{+i(\chi/2)\hat n}\,
    \rho_0\,
    e^{+i(\chi/2)\hat n}\,e^{+i\hat H t}
    \right)\nonumber\\
    &=&
    \mathrm{Tr}\!\left(
    \rho_0\,e^{+i\hat H t}\,e^{+i\chi\hat n}\,e^{-i\hat H t}
    \right)\nonumber\\
    &=&
    \left\langle e^{i\chi \hat n(t)}\right\rangle\,
    \label{zofchi}
\end{eqnarray}
as anticipated. With our choice above, $|\psi(t)\rangle = D(\alpha)|0\rangle$ where $D(\alpha)$ is the displacement operator and $\alpha = -igt$ so 
\begin{eqnarray}
    Z(\chi)&=&\langle 0|D^\dagger(\alpha)\,e^{i\chi \hat n}\,D(\alpha)|0\rangle\nonumber\\
    &=& \langle 0|D^\dagger(\alpha)\,D(e^{i\chi}\alpha)\,e^{i\chi \hat n}|0\rangle\nonumber\\
    &=& \langle 0|D^\dagger(\alpha)\,D(e^{i\chi}\alpha)|0\rangle\,,
\end{eqnarray}
where we have used the displacement operator conjugation $e^{i\chi \hat n}\,D(\alpha)\,e^{-i\chi \hat n}= D(e^{i\chi}\alpha)$ and the fact that $e^{i\chi \hat n}|0\rangle=|0\rangle$ in the last step. Using the Weyl relation
\begin{eqnarray}
    D(\beta)D(\gamma)=e^{\frac{1}{2}(\beta\gamma^*-\beta^*\gamma)}D(\beta+\gamma)\,.
\end{eqnarray}
and the fact that $D^\dagger(\alpha)=D(-\alpha)$ we can write
\begin{eqnarray}
    D^\dagger(\alpha)\,D(e^{i\chi}\alpha)
    &=&
    e^{\frac12\big((- \alpha)(e^{-i\chi}\alpha^*)-(-\alpha^*)(e^{i\chi}\alpha)\big)}
    \,D\big((e^{i\chi}-1)
    \alpha\big)\nonumber\\
    &=& e^{i|\alpha|^2\sin\chi}\,
    D\big((e^{i\chi}-1)\alpha\big)\,,
\end{eqnarray}
which, with a little trigonometry and the standard vacuum overlap $\langle 0|D(\delta)|0\rangle = e^{-\frac12|\delta|^2}$ leads to 
the satisfyingly compact expression
\begin{eqnarray}
    Z(\chi,t)=\exp\!
    \Big(g^2t^2(e^{i\chi}-1)\Big)\,.
    \label{eq:gen-fun-hopping}
\end{eqnarray}
It then follows trivially that
\begin{eqnarray}
    K(t) = \frac{1}{i}\frac{\partial}{\partial\chi}\ln Z(\chi)\Bigg|_{\chi=0}
    = \frac{1}{i}|\alpha|^2(i e^{i\chi})\Bigg|_{\chi=0}=|\alpha|^2 e^{i\chi}\Bigg|_{\chi=0} = g^{2}t^{2}\,.
\end{eqnarray}
This example, like the path-integral computation of the free particle Green's function, might seem like overkill. However, like that example, it both serves as a useful check of the formalism, and does much of the heavy-lifting for the examples to follow.

\subsection{The $\mathfrak{su}(1,1)$ model
}
\noindent
To sharpen the benchmark for our Schwinger–Keldysh formulation, we next focus on a distinguished class of models in which the Liouvillian itself closes into a non-compact Lie algebra. Concretely, we consider systems for which the operator algebra generated by repeated commutation with the Hamiltonian can be organized into a representation of $\mathfrak{su}(1,1)$.
Recall that the Liouvillian $\mathcal L=[H,\cdot]$ acts linearly on the operator Hilbert space $\mathscr H_{\mathrm{op}}$, equipped with the (normalized) Hilbert–Schmidt inner product. In generic interacting systems, the algebra generated by $\mathcal L$ is infinite and structureless. However, in a number of physically important cases including large-$N$ limits, conformal quantum mechanics, and SYK-like models, the Liouvillian algebra closes into a finite set of generators obeying simple commutation relations. We assume that there exist three such operators, $\mathcal L_0$, $\mathcal L_+$, and $\mathcal L_-$,
acting on $\mathscr H_{\mathrm{op}}$, which satisfy the $\mathfrak{su}(1,1)$ commutation relations
\begin{eqnarray}
    [\mathcal L_0,\mathcal L_\pm]=\pm \mathcal L_\pm\,,
    \qquad
    [\mathcal L_+,\mathcal L_-]=-2\mathcal L_0.
    \label{eq:commutator}
\end{eqnarray}
These operators should be thought of as raising, lowering, and Cartan generators in operator space, not in the physical Hilbert space. Importantly, they are constructed from commutators with $H$ and therefore encode operator growth rather than state evolution. Within this algebraic framework, we take the physical Liouvillian governing Heisenberg evolution to be a real linear combination of these generators,
\begin{eqnarray}
    \mathcal L=\alpha(\mathcal L_+ + \mathcal L_-)+\beta \mathcal L_0,
    \label{eq:liouvillian}
\end{eqnarray}
with $\alpha>0$ setting the overall growth scale and $\beta$ allowing for a drift along the Cartan direction. To be concrete (and because this is just a sanity-check), we will focus on the case $\beta=0$. The $\beta\mathcal L_0$ deformation does not modify the qualitative growth structure and can be treated straightforwardly.\\

\noindent
We will assume that the initial operator $|O_0\rangle\in\mathscr H_{\mathrm{op}}$ is a lowest-weight state of the $\mathfrak{su}(1,1)$ representation that satisfies
\begin{eqnarray}
    \mathcal L_-|O_0\rangle=0,
    \quad
    \mathcal L_0|O_0\rangle=k\,|O_0\rangle\,,
    \label{eq:lowestweight}
\end{eqnarray}
where $k>0$ is the Bargmann index characterizing the representation. Physically, this assumption means that the seed operator is ``simple" in the sense that it cannot be further simplified by the Liouvillian lowering operation and is the natural analogue, in operator space, of choosing a lowest-weight state in conformal or algebraic constructions.
Equations \eqref{eq:commutator}, \eqref{eq:liouvillian} and \eqref{eq:lowestweight}, together define precisely the discrete-series representation $D^+(k)$ of $SU(1,1)$. Acting repeatedly with $\mathcal L_+$ generates an infinite orthonormal tower of operators,
\begin{eqnarray}
    |O_n\rangle \;\propto\; (\mathcal L_+)^n |O_0\rangle,
    \qquad n=0,1,2,\dots,
    \label{su11-krylov-basis}
\end{eqnarray}
which we identify with the Krylov basis. In this basis, the Cartan generator acts diagonally, 
\begin{eqnarray}
    \mathcal L_0 |O_n\rangle = (n+k)\,|O_n\rangle,
    \label{su11-cartan}
\end{eqnarray}
while the $\mathcal L_\pm$ act as nearest-neighbor ladder operators with matrix elements fixed entirely by the representation theory. Consequently, the Lanczos coefficients are known exactly,
\begin{eqnarray}
    b_{n+1}
    =
    \alpha\,\sqrt{(n+1)(n+2k)}\,,
    \label{su11-lanczos}
\end{eqnarray}
with $b_0=0$. This, in turn is used to produces closed-form expressions for $\varphi_n(t)$, $P(n,t)$ and $K(t)$. For example,
\begin{eqnarray}
    P(n,t)
    =
    \binom{n+2k-1}{n}
    \frac{\tanh^{2n}(\alpha t)}{\cosh^{4k}(\alpha t)},
    \quad\mathrm{and}\quad
    K(t)=2k\,\sinh^2(\alpha t)\,.
\end{eqnarray}
At large $n$, these coefficients grow linearly, $b_n \sim \alpha n$,
placing this model squarely in the ``maximally chaotic" class according to the usual Krylov criterion. This $SU(1,1)$ construction therefore provides an example of an exactly solvable realization of exponential Krylov growth, with all operator-space dynamics fixed by symmetry. However the construction relies heavily on the existence of the $SU(1,1)$ basis, and hides the dynamical origin of exponential growth.\\

\noindent
Having identified the $\mathfrak{su}(1,1)$ Liouvillian structure underlying the Krylov dynamics, we now compute the Krylov complexity directly using the Schwinger–Keldysh generating functional, keeping in mind that most of the heavy lifting was already covered in the previous example. We start by specializing to the  Liouvillian, $\mathcal L = \alpha(\mathcal L_+ + \mathcal L_-),$
acting on the discrete-series representation $D^+(k)$ with lowest-weight state $|O_0\rangle$. As argued above, the Krylov basis coincides with the weight basis $|n\rangle$, and the position operator is given by $\hat n = \mathcal L_0 - k$. To access the full distribution of $\hat n(t)$, and in particular to compute its moments in a compact way, it is convenient to again introduce a counting field $\chi$ localized at the final time. This is implemented by the symmetric choice
\begin{eqnarray}
    J_+(t')=\frac{\chi}{2}\,\delta(t'-t),
    \quad
    J_-(t')=-\frac{\chi}{2}\,\delta(t'-t)\,,
\end{eqnarray}
and results in the generator \eqref{zofchi} which can now be written purely algebraically as
\begin{eqnarray}
    Z(\chi)
    =
    \langle O_0|
    \,e^{i\mathcal L t}\,
    e^{i\chi(\mathcal L_0-k)}\,
    e^{-i\mathcal L t}
    |O_0\rangle\,.
\end{eqnarray}
The key simplification here is that the time-evolution operator generated by $\mathcal L$ can be disentangled using standard $\mathfrak{su}(1,1)$ identities. Writing
\begin{eqnarray}
    e^{-i\mathcal L t}
    =
    e^{-i\alpha t(\mathcal L_+ + \mathcal L_-)}
    =
    e^{\xi \mathcal L_+ - \xi^* \mathcal L_-},
\end{eqnarray}
with $\xi=-i\alpha t$, we use the Perelomov disentangling formula\footnote{This is the statement that if $G$ is a Lie group with Lie algebra $\mathfrak g$, and 
$X = \sum_a c_a\,T_a \in \mathfrak g$, then 
$e^{X}
=
e^{f_1(X)\,T_{i_1}}
\,e^{f_2(X)\,T_{i_2}}
\cdots
e^{f_k(X)\,T_{i_k}},
$, where the $T_{i_j}$ belong to a chosen ordered basis of $\mathfrak g$, the functions $f_j(X)$ are $c$-number functions determined by the algebra and the ordering reflects a chosen decomposition of the Lie algebra.} to obtain
\begin{eqnarray}
    e^{-i\mathcal L t}
    =
    e^{\tau \mathcal L_+}\,
    e^{\ln(1-|\tau|^2)\mathcal L_0}\,
    e^{-\tau^* \mathcal L_-}\,,
\end{eqnarray}
where $\tau=-i\tanh(\alpha t)$. Acting on the lowest-weight state and using $\mathcal L_-|O_0\rangle=0$, yields
\begin{eqnarray}
    e^{-i\mathcal L t}|O_0\rangle
    =
    (1-|\tau|^2)^k\,e^{\tau \mathcal L_+}|O_0\rangle
    \equiv |\tau\rangle\,,
\end{eqnarray}
where $|\tau\rangle$ is an $SU(1,1)$ coherent state in operator space. The generating function therefore reduces to the expectation value of $e^{i\chi\hat n}$ in this coherent state,
\begin{eqnarray}
   Z(\chi)=\langle \tau|e^{i\chi(\mathcal L_0-k)}|\tau\rangle\,. 
\end{eqnarray}
Using the adjoint action
\begin{eqnarray}
    e^{i\chi\mathcal L_0}\,\mathcal L_+\,e^{-i\chi\mathcal L_0}
    =
    e^{i\chi}\mathcal L_+\,,
\end{eqnarray}
together with the defining properties of the discrete series, we finds after a short calculation
\begin{eqnarray}
    Z(\chi,t)
    =
    \left(
    \frac{1-|\tau|^2}{1-e^{i\chi}|\tau|^2}
    \right)^{2k}\,,
    \label{su11-full-gen-fun}
\end{eqnarray}
where $|\tau|^2=\tanh^2(\alpha t)$. Note that This expression is obtained entirely within the Schwinger-Keldysh framework and makes no reference to an explicit solution of the Krylov recursion or to known probability distributions. We can then compute the Krylov complexity immediately from the cumulant-generating function,
\begin{eqnarray}
    \ln Z(\chi)
    =
    2k\Big[
    \ln(1-|\tau|^2)
    -
    \ln(1-e^{i\chi}|\tau|^2)
    \Big]\,,
\end{eqnarray}
by differentiation,
\begin{eqnarray}
   K(t)&=&
   \left.
   \frac{1}{i}\,\partial_\chi \ln Z(\chi)
   \right|_{\chi=0}\nonumber\\
   &=& 2k\,\frac{|\tau|^2}{1-|\tau|^2}
   = 2k\,\sinh^2(\alpha t)\,.
\end{eqnarray}
Higher cumulants are equally straightforward to derive. At late times, $K(t)\sim \frac{k}{2}\,e^{2\alpha t}$, exhibits exponential Krylov growth with growth rate $2\alpha$. In our Schwinger-Keldysh formulation however, this growth arises from the instability encoded in the non-compact $SU(1,1)$ structure of the Liouvillian and, as we show below, can be reinterpreted semiclassically as motion along hyperbolic trajectories in Krylov phase space.\\

\noindent
We note here that not only do the generating functionals  constructed in \eqref{eq:gen-fun-hopping} and \eqref{su11-full-gen-fun} determine the mean Krylov complexity, they actually encode the complete probability distribution of the
Krylov position and, by extension, the full set of higher cumulants (see \eqref{eq:cumulants} below). For the hopping case, for example, we find that $\kappa_m(t)=g^2t^2$ and consequently that $\kappa_2/K^2 = 1/g^2t^2 \to 0$, so
that the Krylov distribution becomes self-averaging at late times. On the other hand, for the SU(1,1) chain, \eqref{su11-full-gen-fun} gives $\kappa_2 = K+\frac{K^2}{2k}$, and hence $\frac{\kappa_2}{K^2}\rightarrow\frac{1}{2k}$. As a result, the hyperbolically growing Krylov distribution retains finite relative fluctuations, whereas the square-root chain becomes increasingly sharply concentrated about its mean. This distinction is invisible at the level of the mean complexity alone. 


\section{Semiclassical Limit and Hyperbolic Trajectories}
\label{semiclassics}
\noindent
In this section we derive the semiclassical dynamics of Krylov complexity directly from the Schwinger–Keldysh phase-space action and show how exponential growth arises from hyperbolic classical trajectories. This provides a dynamical explanation of the empirical observation that linear Lanczos growth is associated with chaotic operator dynamics.\\

\noindent
Starting from the Schwinger–Keldysh generating functional for the Krylov chain,
\begin{eqnarray}
    Z[J_+,J_-]
    =
    \int \mathcal{D}n_\pm \mathcal{D}p_\pm\;
    \exp\Bigl(
    iS[n_+,p_+;J_+]
    -
    iS[n_-,p_-;J_-]
    \Bigr),
\end{eqnarray}
with branch action
\begin{eqnarray}
    S[n,p;J]
    =
    \int_0^t dt'\,
    \Big(
    p\,\dot n
    -
    H_{\mathrm{eff}}(n,p)
    +
    J(t')\,n
    \Big)\,,
\end{eqnarray}
if we introduce the ``classical" and ``quantum| fields, 
\begin{eqnarray}
    n_c=\frac{n_++n_-}{2},\quad n_q=n_+-n_-,
    \quad
    p_c=\frac{p_++p_-}{2},\quad p_q=p_+-p_-,
\end{eqnarray}
then the Keldysh action takes the generic form
\begin{eqnarray}
    S_K
    &=&
    \int_0^t dt\,
    \Big[
    p_q \dot n_c + p_c \dot n_q
    -
    H_{\mathrm{eff}}(n_c+\tfrac12 n_q,p_c+\tfrac12 p_q)\nonumber\\
    &+&
    H_{\mathrm{eff}}(n_c-\tfrac12 n_q,p_c-\tfrac12 p_q)
    +
    J_q n_c + J_c n_q
    \Big]\,.
\end{eqnarray}
In the unitary (closed-system) case, the action is linear in the quantum fields at leading order. Varying with respect to $n_q$ and $p_q$ yields the saddle-point equations
\begin{eqnarray}
    \dot n_c = \partial_p H_{\mathrm{eff}}(n_c,p_c),
    \qquad
    \dot p_c = -\partial_n H_{\mathrm{eff}}(n_c,p_c),    \label{classicalSKequatoins}
\end{eqnarray}
which are just Hamilton’s equations generated by $H_{\mathrm{eff}}(n,p)$. In other words, in the semiclassical limit, Krylov dynamics reduces to classical motion in an emergent phase space with canonical coordinates (n,p).  \\ \\
Before proceeding, a comment is in order.  The equations (\ref{classicalSKequatoins}) are first order and share some similarities with those obtained obtained when considering the continuum limit in \cite{Muck:2022xfc, Erdmenger:2023wjg}.  The continuum limit relies on the fact that all quantities involved are sufficiently smooth functions of $x$.  This is not the case for the Krylov amplitude function ($\phi_n(t) \rightarrow \phi(x, t)$ ) at $t=0$.  Due to this, the limit is usually considered to hold after some finite time has elapsed.  However, the function $n(t)$ is smooth for all values of $t$ and, as such, we expect a sensible continuum limit for all $t \geq 0$. 

\subsection{Hyperbolic trajectories}
We now specialize to the regime of interest for chaotic systems, in which the Lanczos coefficients grow linearly at large $n$,
$b(n)\sim \alpha n$, for $n\gg 1$. To leading order, the Weyl symbol of the hopping Hamiltonian reduces to
\begin{eqnarray}
    H_{\mathrm{eff}}(n,p)
    =
    2 b(n)\cos p
    \simeq
    2\alpha n\cos p.
\end{eqnarray}
Corrections arising from gradients of $b(n)$ are subleading at large $n$ and will be discussed below. The classical equations of motion are therefore
\begin{eqnarray}
    \dot n = -2\alpha n\sin p,
    \qquad
    \dot p = -2\alpha\cos p.
\end{eqnarray}
This system is autonomous and integrable, but it exhibits a nontrivial phase-space structure, including hyperbolic fixed points.
More precisely, the phase portrait of the system in Figure. 1 reveals two special lines at $p=\pm\frac{\pi}{2}$. Along these lines,
$\dot p = 0$, and $\dot n = \mp 2\alpha n$.
Consequently,
\begin{eqnarray}
    n(t) \propto e^{\pm 2\alpha t},
\end{eqnarray}
corresponding to exponentially growing or decaying trajectories in Krylov space. The growing branch at $p=-\pi/2$ is dynamically attractive, while the decaying branch is repulsive.\\

\noindent
Linearizing around the growing trajectory,
$p(t) = -\frac{\pi}{2} + \delta p(t)$,
it is clear that
$\dot{\delta p} = -2\alpha\,\delta p$,
confirming that fluctuations transverse to the hyperbolic trajectory decay exponentially. This stability explains why generic initial conditions evolve toward the exponentially growing solution at late times.
The exponential growth rate
$\lambda_K = 2\alpha$ can therefore be identified as a Lyapunov exponent in Krylov phase space. This provides a precise \textit{dynamical} interpretation of exponential Krylov complexity growth as a manifestation of classical instability.

\begin{figure}
\centering
\includegraphics[width=0.5\textwidth]{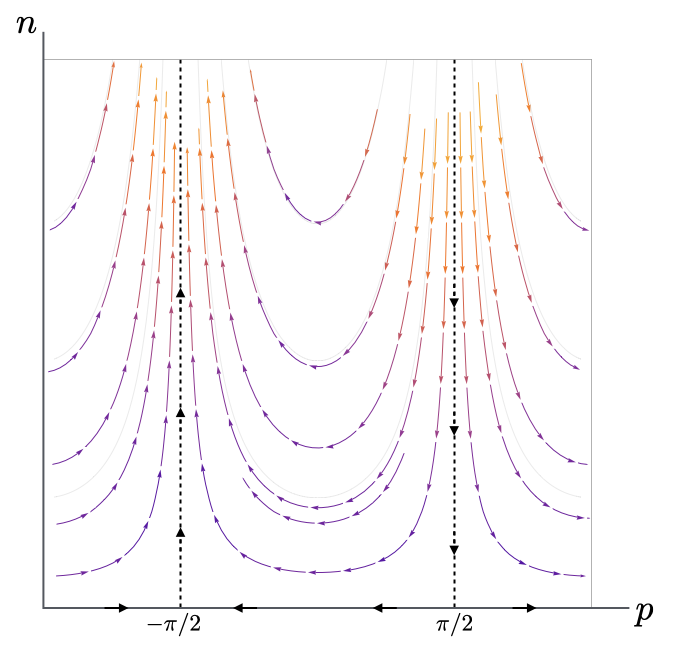}
\caption{Phase portrait for the semiclassical Krylov dynamics
$\dot n=-2\alpha n\sin p$, $\dot p=-2\alpha\cos p$.
Trajectories lie on $n\cos p=\mathrm{const}$.
The vertical manifolds at $p=\pm\pi/2$ are fixed in $p$; $p=-\pi/2$ is unstable (repels in $n$) and
$p=+\pi/2$ is stable (attracts in $n$).}
\label{Krylovphase}
\end{figure}

\subsection{Linear growth as a universal fixed point}
\noindent
In the semiclassical Keldysh phase space, the effective classical Hamiltonian is $H_{\rm eff}(n,p)=2\,b(n)\cos p$,
so
\begin{eqnarray}
    \dot n=\partial_p H_{\rm eff}=-2\,b(n)\sin p,
    \qquad
    \dot p=-\partial_n H_{\rm eff}=-2\,b'(n)\cos p.
\end{eqnarray}
The ``chaotic fixed point" is
\begin{eqnarray}
   b(n)=\alpha n\quad\Rightarrow\quad b'(n)=\alpha\,. 
\end{eqnarray}
Then $\dot p\propto \cos p$ is $n$-independent, and the flow has noncompact hyperbolic manifolds at $\cos p=0$, \textit{i.e.} $p=\pm\pi/2$. Linearizing about the unstable one, we find that
\begin{eqnarray}
    p(t)= -\frac{\pi}{2}+\delta p(t)\,,
\end{eqnarray}
with $|\delta p|\ll1$, gives $\cos p\simeq \delta p$ and $\sin p\simeq -1,$ hence
\begin{eqnarray}
    \dot n \simeq 2 b(n) \simeq -2\alpha n,
    \qquad
    \dot{\delta p}\simeq -2\alpha\,\delta p,
\end{eqnarray}
so one direction contracts and the conjugate expands corresponding to a genuine hyperbolic saddle structure. This is the geometric origin of exponential growth and the statement that ``linear $b_n$ implies chaos".
Now let's perturb this fixed point by writing
\begin{eqnarray}
    b(n)=\alpha n + \delta b(n),
    \qquad
    \delta b(n)\ll \alpha n \ \text{for large }n,
\end{eqnarray}
and ask whether $\delta b$ changes the existence (or even just the strength) of the hyperbolic instability. A simple but systematic way to organize this is to classify $\delta b(n)$ by its scaling at large $n$, $\delta b(n)\sim n^\gamma$
with $\gamma<1$, $=1$ or $>1$, or by softer behaviors like $\log n$, constants or even oscillations. The hyperbolic mechanism depends on two ingredients; the large-$n$ drift through $\dot n \sim b(n)$ near $\sin p\simeq\pm1$, and the angular instability controlled by $b'(n)$ through $\dot p \sim b'(n)\cos p$. As a result, the ``universality class" is set by the asymptotics of both $b(n)$ and $b'(n)$. Borrowing terminology\footnote{We stress here that this terminology is meant simply as an analogy for the asymptotic classification of deformations of the Lanczos coeﬃcients according to their influence on long-time semiclassical dynamics with no claim of any Wilsonian RG construction involving explicit coarse-graining, beta functions, or RG trajectories.} from renormalization group theory we organize the perturbations as follows:
\begin{itemize}
    \item \textbf{Irrelevant perturbations:}    The simplest and most instructive class of deformations consists of constant shifts in the Lanczos coefficients,
    \begin{eqnarray}
        b(n)=\alpha(n+c)
        \qquad\text{or equivalently}\qquad
        b(n)=\alpha n+\gamma\,,
    \end{eqnarray}
    with $\gamma=\alpha c$, possibly supplemented by corrections that decay at large $n$ and which are $O(1/n)$. Such terms naturally arise from microscopic details of the Hamiltonian or from the choice of seed operator, and they are more-or-less ubiquitous in exactly solvable models. From the semiclassical point of view, the effect of this deformation is immediately transparent. The effective Hamiltonian governing Krylov phase-space dynamics,
    $H_{\rm eff}(n,p)=2\,b(n)\cos p$,
    now reads $H_{\rm eff}(n,p)=2\alpha(n+c)\cos p$.
    Crucially, the derivative that controls the angular flow, $b'(n)=\alpha$,
    is unchanged. As a result, the structure of the phase portrait near the noncompact fixed manifolds $p=\pm\pi/2$ is identical to that of the strict fixed point $b(n)=\alpha n$. The locations, stability properties, and hyperbolic nature of these manifolds are unaffected; only the origin of the $n$-coordinate is shifted.\\

    \noindent
    This implies that the mechanism responsible for exponential Krylov growth survives intact. Near the unstable manifold at $p=-\pi/2$, where $\sin p\simeq-1$, the radial equation of motion becomes $\dot n \simeq 2\alpha(n+c)$.
    Solving this yields $n(t)=(n_0+c)\,e^{2\alpha t}-c$,
     so that at late times
    $K(t)=\langle n(t)\rangle \;\propto\; e^{2\alpha t}$, as $t\to\infty$.
    The growth rate $2\alpha$ remains  unchanged with only the prefactor and the early-time crossover depending on the shift $c$. The deformation modifies the ultraviolet structure of the Krylov chain in how the dynamics behaves at small $n$, but leaves the infrared instability that controls late-time operator growth untouched. A concrete realization of this structure is provided by the $SU(1,1)$ discrete-series representations discussed above. In that case the Lanczos coefficients are known exactly, 
    \begin{eqnarray}
        b_{n+1} = \alpha\sqrt{(n+1)(n+2k)} = \alpha\left(n+k+\mathcal O(1/n)\right)\,,
    \end{eqnarray}
    where $k>0$ is the Bargmann index labeling the representation. From the semiclassical perspective, the entire one-parameter family of $SU(1,1)$ models differs from the linear fixed point only by the constant shift $c=k$. Changing $k$ alters the normalization of the coherent-state orbit and the short-time behavior of $K(t)$, but it does not modify the hyperbolic instability or the exponential growth rate. In other words, the Bargmann index $k$ labels different ultraviolet completions of the same infrared Krylov universality class. Models with different microscopic operator content or seed operators may correspond to different values of $k$, yet they all flow to the same semiclassical fixed point characterized by linear $b(n)$ and exponential Krylov growth. From this viewpoint, the well-known robustness of exponential complexity growth in $SU(1,1)$-type systems is precisely the geometric manifestation of irrelevance in Krylov phase space.
    
    \item  \textbf{Irrelevant perturbations with slow drift:} In many physically relevant situations the approach to the linear fixed point can be logarithmically slow. This is precisely what is expected in nearly conformal dynamics where irrelevant operators induce slow running. In the Krylov language, this corresponds to Lanczos coefficients of the form $b(n)=\alpha n+\beta\ln n$, with $n\gg1$ and $|\beta|\ll \alpha n$ at large $n$. The key point is that this deformation remains asymptotically linear; it therefore does not destroy the hyperbolic mechanism responsible for exponential complexity growth, but it does imprint universal subleading corrections. For this deformation,
    \begin{eqnarray}
        b'(n)=\alpha+\frac{\beta}{n} \to \alpha\,,
    \end{eqnarray}
    as $n\to\infty$, and the angular flow near the fixed manifolds $p=\pm\pi/2$ is asymptotically unchanged. In particular, the noncompact manifolds at $\cos p=0$ remain fixed in $p$, and the local hyperbolic structure persists with the same asymptotic instability scale set by $\alpha$. In RG terms, the chaotic fixed point survives and the logarithmic correction induces only a slow drift in the radial direction. To extract the leading correction to the growth of $n(t)$, it suffices to examine motion near the growth manifold $p\simeq -\pi/2$,
    where the radial equation becomes
    \begin{eqnarray}
        \dot n \simeq -2\,b(n)\sin p
        \simeq 2\,b(n)
        \simeq 2\alpha n+2\beta\ln n\,.
        \label{radial-eq}
    \end{eqnarray}
    A convenient way to solve this asymptotically is to factor out the dominant exponential growth by writing
    $n(t)=e^{2\alpha t}\,y(t)$. Substituting into \eqref{radial-eq} gives
    \begin{eqnarray}
        \dot y(t)
        =
        2\beta\,e^{-2\alpha t}\,
        \ln\!\big(e^{2\alpha t}y(t)\big)
        =
        2\beta\,e^{-2\alpha t}\,
        \big(2\alpha t+\ln y(t)\big).
    \end{eqnarray}
    At late times, the exponential suppression $e^{-2\alpha t}$ ensures that $y(t)$ varies slowly, and may be approximated self-consistently by $y(t)\approx n_0$ inside the logarithm. Integrating this to leading order then gives
    \begin{eqnarray}
        y(t)
        &\simeq&
        n_0
        +
        2\beta\int_0^t ds\,e^{-2\alpha s}\big(2\alpha s+\ln n_0\big)\nonumber\\
        &\simeq&
        n_0
        +
        \frac{\beta}{\alpha}\left[
        1-e^{-2\alpha t}
        +
        \ln n_0\big(1-e^{-2\alpha t}\big)
        -2\alpha t\,e^{-2\alpha t}
        \right]\,.
    \end{eqnarray}
    Substituting back into the definition gives,
    \begin{eqnarray}
        n(t) \simeq
        n_0 e^{2\alpha t}
        +
        \frac{\beta}{\alpha}e^{2\alpha t}(1+\ln n_0)
        -\frac{\beta}{\alpha}(1+\ln n_0)
       -2\beta t
       +\cdots.
    \end{eqnarray}
    The dominant behavior remains exponential with rate $2\alpha$. The logarithmic deformation produces a controlled multiplicative renormalization of the amplitude, together with subleading additive corrections that grow only linearly in time. Specifically as $t\to\infty$,
    \begin{eqnarray}
        n(t)\sim
        \Big(n_0+\frac{\beta}{\alpha}(1+\ln n_0)\Big)e^{2\alpha t}\,,
    \end{eqnarray}
    up to corrections parametrically smaller than $e^{2\alpha t}$. Continuing with our RG analogy, the $\beta\log n$ deformation is therefore irrelevant. It ``renormalizes" nonuniversal prefactors without modifying the universal growth exponent. While both perturbations are irrelevant, the logarithmic term introduces a slow running at intermediate scales. In the phase-space picture, this manifests as a mild $n$-dependence of the angular flow through
    $\dot p=-2\big(\alpha+\beta/n\big)\cos p$.
    The hyperbolic manifolds persist, but the approach to the asymptotic regime acquires a weak scale dependence. Such subleading structure is difficult to characterize using continued fractions alone. Logarithmic corrections of the above form arise whenever an emergent conformal or $SU(1,1)$-like operator-growth structure is softly broken by irrelevant perturbations such as nearly conformal regimes where a Schwarzian sector is perturbed by higher-dimension operators, as well as SYK-like dynamics away from the strict conformal window. While the coefficient $\beta$ is model-dependent, the universality classification is not. The defining property is the asymptotic approach $b'(n)\to\alpha$, which guarantees the persistence of the hyperbolic growth mechanism and fixes the universal exponent $2\alpha$.
    
    \item \textbf{Marginal perturbations:} A qualitatively different class of deformations are of the form,
    \begin{eqnarray}
        b(n)=\alpha n\left(1+\frac{\epsilon}{\ln n}\right)\,,
    \end{eqnarray}
    with $n\gg 1$ and $|\epsilon|\ll \ln n$,
    sit precisely at the boundary between relevance and irrelevance. These perturbations modify the \textit{slope} of the Lanczos coefficients in a way that vanishes only logarithmically at large $n$. Such terms are therefore marginal in that they do not destroy the hyperbolic instability responsible for exponential growth, but they induce slow, universal running of the effective instability exponent. To see this explicitly, note that near the unstable manifold $p\simeq-\pi/2$, the radial equation reduces to
    \begin{eqnarray}
        \dot n \simeq 2\,b(n)
        \simeq
        2\alpha n\left(1+\frac{\epsilon}{\ln n}\right)\,.
    \end{eqnarray}
    This can equivalently be rewritten as
    \begin{eqnarray}
        \frac{d}{dt}\ln n
        =
        \frac{\dot n}{n}
        \simeq
        2\alpha\left(1+\frac{\epsilon}{\ln n}\right)\,,
        \label{logn}
    \end{eqnarray}
    which, at leading order gives the familiar exponential growth, but the subleading term introduces a slow running of the effective growth exponent,
    \begin{eqnarray}
        \lambda_{\rm eff}(n)
        \equiv
        2b'(n)
        \simeq
        2\alpha\left(1+\frac{\epsilon}{\ln n}\right).
    \end{eqnarray}
    Defining $u(t):=\ln n$ allows us to write \eqref{logn} as
    \begin{eqnarray}
        \dot u = 2\alpha\left(1+\frac{\epsilon}{u}\right)
        \quad\Rightarrow\quad
        dt = \frac{u}{2\alpha(u+\epsilon)}\,du\,,
    \end{eqnarray}
    which can be integrated to get,
    \begin{eqnarray}
        t &=& \frac{1}{2\alpha}\int \frac{u}{u+\epsilon}\,du
        = \frac{1}{2\alpha}\int\left(1-\frac{\epsilon}{u+\epsilon}\right)du\nonumber\\
        &=& \frac{1}{2\alpha}\left(u-\epsilon\ln(u+\epsilon)\right)+\text{const}\,,
    \end{eqnarray}
    or, equivalently,
    \begin{eqnarray}
        u - \epsilon\ln(u+\epsilon) = 2\alpha t + \text{const}.
    \end{eqnarray}
    Then, using the fact that at large $t$, $u(t)\sim 2\alpha t$ is large, we can replace $\ln(u+\epsilon)=\ln u +O(1/u)$ and $\ln u = \ln(2\alpha t)+O(\ln\ln t)$ to write
    \begin{eqnarray}
        u = 2\alpha t + \epsilon\ln t + O(1)\,,
    \end{eqnarray}
    where the $O(1)$ absorbs $\epsilon\ln(2\alpha)$, constants from initial data, and the tiny $\ln\ln t$ terms. Finally, exponentiating gives
    \begin{eqnarray}
        n(t) = e^{u(t)} = Ae^{2\alpha t}t^{\epsilon}\,,
    \end{eqnarray}
    where $A$ is some non-universal constant depending on the initial conditions. Clearly, the exponential growth rate $2\alpha$ remains intact, but is now dressed by a universal power-law correction. This behavior is the defining signature of a marginal deformation where the system neither flows away from the chaotic fixed point nor remains strictly scale invariant, but instead exhibits logarithmic running.
    From the phase-space perspective, the geometry of the flow remains hyperbolic, and the angular velocity $\dot p\sim b'(n)\cos p$ acquires a slow dependence on $n$. Trajectories are still attracted toward and repelled from the same noncompact manifolds, but the rate at which they approach the asymptotic regime drifts logarithmically. This drift should accumulate over long times and leave an observable imprint on $K(t)$. From the Krylov standpoint, this is precisely the regime in which continued-fraction or Lanczos methods become cumbersome in the sense that the asymptotic slope is well defined, but extracting the logarithmic corrections requires detailed control over many Lanczos coefficients. In our semiclassical Schwinger–Keldysh formulation, marginality manifests directly as a scale-dependent instability exponent.

    \item \textbf{Relevant perturbations:} There is a class of perturbations that qualitatively alter the semiclassical Krylov dynamics and destroy the hyperbolic instability responsible for exponential complexity growth corresponding to sublinear growth of the Lanczos coefficients,
    \begin{eqnarray}
        b(n)\sim n^\gamma,
        \qquad 0<\gamma<1,
        \label{relevant}
    \end{eqnarray}
    which is parametrically slower than the linear fixed point $b(n)=\alpha n$. Such perturbations are relevant in the sense that they grow and drive the system away from the chaotic universality class. To appreciate the implications of \eqref{relevant}, note that for this perturbation,
    \begin{eqnarray}
        b'(n)\sim n^{\gamma-1}
        \;\xrightarrow[n\to\infty]{}\;0\,,
    \end{eqnarray}
    which means that as $n$ becomes large, the angular velocity $\dot p$ vanishes and the flow in the $p$-direction freezes asymptotically. This should be contrasted with the chaotic fixed point, where $b'(n)\to\alpha$ produces a finite angular velocity that dynamically attracts trajectories toward the noncompact hyperbolic manifolds at $p=\pm\pi/2$. As a result, the system is no longer forced into the hyperbolic sector of phase space. Explicitly, near 
    $p\simeq -\frac{\pi}{2}$, where
    $\sin p\simeq -1$, the radial equation reduces to $\dot n = -2\,b(n)\sin p \simeq 2\,n^\gamma$ which can be integrated to give
    \begin{eqnarray}
        \int n^{-\gamma}\,dn \sim 2\int dt
        \quad\Rightarrow\quad
        n(t)\sim t^{\frac{1}{1-\gamma}},
    \end{eqnarray}
    up to multiplicative constants set by initial conditions. The growth of $n(t)$ is therefore polynomial rather than exponential. Equivalently, the effective instability exponent vanishes at late times, in sharp contrast with the marginal and irrelevant cases.
\end{itemize}

\section{Krylov Trajectories and Phase Transitions in Operator Growth}
\noindent
A central lesson of the Schwinger–Keldysh formulation developed in this work is that Krylov complexity is a statistical object describing an ensemble of operator-growth trajectories in Krylov space. This perspective naturally raises the question of how operator growth behaves across phase transitions, and which transitions can be meaningfully detected by Krylov-based observables. The key point is that the generating functional
$Z(\chi,t)=\left\langle e^{\,i\chi \hat n(t)}\right\rangle$
defines a \textit{dynamical free energy} for operator growth trajectories,
\begin{eqnarray}
    \Psi(\chi)     =
    \lim_{t\to\infty}\frac{1}{t}\ln Z(\chi,t)\,,
\end{eqnarray}
whose non-analyticities signal trajectory-level phase transitions. This in turn provides a finer-grained probe of dynamical crossovers than the mean complexity $K(t)=\langle \hat n(t)\rangle$ alone. These transitions need not coincide with equilibrium thermodynamic transitions, nor with sharp changes in spectral statistics or Lyapunov exponents. Instead, they reflect qualitative changes in the structure of operator-growth histories themselves.\\

\noindent
To illustrate this, consider a generic situation in which operator growth interpolates between an integrable-like regime and a chaotic regime as a control parameter $h$ is tuned. In Krylov language, this crossover is encoded in the behavior of the Lanczos coefficients $b_n(h)$. Motivated by both numerical studies and general locality arguments, we assume that $b_n(h)$ approaches the linear chaotic fixed point only beyond a parametrically large scale $n_\ast(h)$, which diverges as $h\to 0$. A convenient coarse-grained parametrization is
\begin{equation}
  b(n;h) =
  \alpha(h)\,n\,f\!\left(\frac{n}{n_\ast(h)}\right)
  +
  \gamma(h),
  \label{eq:bn-crossover}
\end{equation}
where $f(x)\to 0$ for $x\ll1$ and $f(x)\to 1$ for $x\gg1$. The detailed form of $f(x)$ is not important but a simple two-parameter choice with the appropriate behavior is\footnote{ $f(x) = x/(1+x)$ can be swapped out for $\tanh x$ or $1-e^{-x}$ without affecting the analysis qualitatively.}
\begin{eqnarray}
    f(x)=\frac{x}{1+x}
    \quad\Rightarrow\quad
    b(n)=\alpha\,n\frac{n}{n+n_\ast}+\gamma.
\end{eqnarray}
Note that for $n\ll n_\ast$,  $b(n)\sim \alpha n^2/n_\ast$ which corresponds to very weak growth, while for $n\gg n_\ast$, $b(n)\sim \alpha n + \gamma$.\\

\noindent
The mean Krylov complexity is governed by the \textit{typical semiclassical trajectory} of the Schwinger-Keldysh action. For any fixed $h\neq 0$, this trajectory eventually reaches the asymptotically linear regime $b(n)\simeq \alpha(h)n$, where the effective phase-space flow becomes hyperbolic. As a result, the late-time growth of $K(t)$ is exponential,
\begin{equation}
    K(t)\sim e^{2\alpha(h)t},
\end{equation}
with a rate that varies smoothly with $h$. The integrability–chaos crossover therefore leaves no sharp signature in the mean complexity alone, consistent with existing numerical observations. The situation is qualitatively different for \textit{fluctuations} of the Krylov position. The full counting statistics
encoded in $Z(\chi)$ and $\Psi(\chi)$ define a hierarchy of cumulants 
\begin{eqnarray}
    \kappa_m(t)
    \;=\;
    \left.
    \frac{\partial^m}{\partial (i\chi)^m}
    \ln Z(\chi,t)
    \right|_{\chi=0}\,,
    \label{eq:cumulants}
\end{eqnarray}
and, in the long-time limit, a large-deviation rate function\footnote{In non-equilibrium systems, the large-deviation rate function $\Phi(v)$ is the Legendre–Fenchel transform of the scaled cumulant generating function.} $\Phi(v) = \sup_{\chi\in\mathbb R}
\big[\,i\chi\,v - \Psi(\chi)\big]$. In the Schwinger-Keldysh formulation, these quantities are controlled by fluctuations around the semiclassical saddle and by the stability properties of the associated phase-space flow. For the crossover form \eqref{eq:bn-crossover}, the system spends a long time in a weakly hyperbolic regime $n\lesssim n_\ast(h)$, where $b'(n)\ll \alpha(h)$ and angular motion in phase space is slow. Only after reaching $n\sim n_\ast(h)$ does the trajectory enter the fully hyperbolic region responsible for exponential growth. The corresponding escape-time scales as
\begin{equation}
t_\ast(h)\sim \frac{1}{2\alpha(h)}\log n_\ast(h),
\end{equation}
which diverges as $h\to 0$. Because different trajectories linger for different durations in this weakly hyperbolic regime, fluctuations in the Krylov position are strongly enhanced. In particular, the second cumulant per unit time,
\begin{equation}
\chi_K
\equiv
\lim_{t\to\infty}\frac{1}{t}\,\mathrm{Var}[n(t)]
=
\left.\frac{\partial^2\psi}{\partial (i\chi)^2}\right|_{\chi=0},
\end{equation}
is controlled by the inverse stability operator around the saddle and is therefore sensitive to $t_\ast(h)$. The Schwinger-Keldysh loop expansion (see Appendix B for details) predicts that $\chi_K$ grows parametrically with the crossover scale, schematically
\begin{equation}
   \chi_K(h)\;\propto\; t_\ast(h)\;\sim\;\frac{1}{2\alpha(h)}\log n_\ast(h),
\end{equation}
as $h\to 0$, up to nonuniversal prefactors.\\

\noindent
The key point is that the integrability–chaos crossover can be smooth in the mean but sharp in fluctuations. While $K(t)$ probes only the typical operator-growth trajectory, higher cumulants and large-deviation functions probe the full ensemble of trajectories and are therefore sensitive to rare events and long dwell times in quasi-integrable regions of phase space. This mechanism is universal in the sense that it depends only on the existence of a parametrically large crossover scale $n_\ast(h)$, not on the detailed form of the interpolation in \eqref{eq:bn-crossover}. From this perspective, the Schwinger-Keldysh formulation raises Krylov complexity from a single diagnostic to a dynamical fluctuation theory, capable of detecting sharp crossovers in operator growth even when no thermodynamic or spectral phase transition is present.


\section{Discussion}
\noindent
In this work we have developed a real-time, Schwinger–Keldysh formulation of Krylov complexity that elevates the standard Lanczos construction into an effective \textit{dynamical} theory of operator growth. By coupling sources to the Krylov position operator and working on a closed time contour, we constructed a generating functional that naturally encodes the full distribution of operator spreading along the Krylov chain. This framework makes explicit several structures that are implicit, but largely hidden, in the recursion-based approach.\\

\noindent
Conceptually, the key step is to recognize Krylov complexity as an in–in observable afterwhich the Schwinger–Keldysh contour becomes unavoidable, and the resulting phase-space path integral follows in a controlled and essentially unique manner. In this formulation, we have showed that the Lanczos coefficients define an emergent classical Hamiltonian governing motion in Krylov phase space. Exponential complexity growth then acquires a clear geometric interpretation as arising from hyperbolic classical trajectories, with the growth rate set by the instability exponent of the phase space flow. In this sense, linear growth of the Lanczos coefficients appears not just as an empirical diagnostic of chaos, but as a genuine fixed point of the effective semiclassical dynamics. One immediate payoff of this reorganization is a sharpened notion of universality. Our semiclassical analysis shows that the long-time behavior of Krylov complexity depends only on the asymptotic scaling of the Lanczos coefficients and their derivatives. Constant shifts, logarithmic corrections, and other subleading deformations are analogous to irrelevant or marginal perturbations in the sense of an effective renormalization group acting on Krylov phase space. Sublinear growth of the Lanczos coefficients, in contrast, are analogous to a relevant deformation that destroys the hyperbolic structure and replaces exponential  by polynomial growth. Framed this way, various disparate results in the Krylov literature can be understood as manifestations of a small number of universality classes, rather than as isolated model-specific observations. \\

\noindent
A second advantage of our Schwinger–Keldysh formulation is that it provides direct access to fluctuations and rare events. The generating functional naturally produces not only the mean complexity $K(t)$, but also higher cumulants and the full counting statistics of the Krylov position. Indeed, by performing a systematic loop expansion around the semiclassical saddle, we showed that fluctuations are governed by the inverse stability operator of the classical flow. This leads us to conjecture that \textit{near integrability–chaos crossovers, where the semiclassical dynamics is only weakly hyperbolic over a long crossover scale, Krylov fluctuations are parametrically enhanced even when the mean complexity varies smoothly}. From this perspective, quantities such as the variance or the large-deviation rate function play a role analogous to susceptibilities in equilibrium statistical mechanics, providing a finer diagnostic of dynamical crossovers than the mean alone. The path-integral formulation also clarifies the status of exactly solvable models. In the square-root hopping example and the $SU(1,1)$ Liouvillian models, the Schwinger–Keldysh machinery reproduces known results for $K(t)$ and the full distribution $P(n,t)$, while making transparent the underlying dynamical mechanism. In particular, the $SU(1,1)$ discrete series emerges as a concrete realization of the linear-growth fixed point, with the Bargmann index labeling different ultraviolet completions of the same infrared universality class. The algebraic solvability of these models has long been appreciated in the context of conformal quantum mechanics and SYK-like systems, but the present formulation shows how these structures fit naturally into a broader dynamical framework for operator growth.\\

\noindent
It is useful, in this regard, to distinguish the different roles played by the examples and results developed above. Several of the growth laws and exactly solvable Krylov chains considered in this work are established results, and their reproduction here should be viewed as benchmarks of the Schwinger–Keldysh construction rather than as new predictions. What the present formulation adds in these cases is a new semiclassical interpretation where the asymptotic Lanczos data define an emergent phase-space geometry, and familiar regimes of Krylov growth correspond to qualitatively distinct classical flows within that geometry. Beyond this reinterpretation, however, the generating functional contains genuinely new information that is absent from the mean complexity alone. In particular, it gives direct access to the hierarchy of cumulants allowing us to distinguish dynamical regimes according to their fluctuations and degree of self-averaging. The loop expansion further provides a systematic framework for computing corrections about the semiclassical limit and relating these fluctuations to the stability properties of the underlying classical flow. We see the role of the Schwinger–Keldysh formulation as threefold. It reproduces known Krylov results as consistency checks, supplies a geometric and semiclassical organization of those results, and provides new observables and controlled extensions that are difficult to access within the recursion formulation alone.\\

\noindent
Several open problems remain. One natural extension of this work is a more detailed exploration of dynamical phase transitions in operator growth, particularly in systems without sharp thermodynamic transitions. The large-deviation structure uncovered here suggests that crossovers between integrable and chaotic regimes may leave sharper signatures in fluctuation diagnostics than in the mean complexity itself. Another direction concerns finite-size effects and saturation. While the semiclassical picture captures the approach to exponential growth, understanding how reflections from the ultraviolet boundary of the Krylov chain feed back into the dynamics will require going beyond leading order and incorporating boundary terms more carefully. Finally, it would be interesting to connect the present formulation to other notions of complexity and scrambling, including circuit complexity and entanglement growth, to clarify which features are universal across different measures and which are genuinely Krylov-specific.\\

\noindent
More broadly, this work reframes Krylov complexity as a problem in non-equilibrium dynamical field theory. By bringing Schwinger–Keldysh methods into the study of operator growth, it opens a route to applying a wide range of tools such as semiclassical analysis, fluctuation theory, and large deviations, that are standard elsewhere in quantum many-body physics but have so far played little role in the Krylov context. We hope that this perspective will help organize existing results and stimulate new applications of Krylov complexity across quantum chaos, thermalization, and open-system dynamics.

\section*{Note Added in Proof} Since the completion of this manuscript, the extension of the Schwinger–Keldysh framework developed here to open quantum systems has been carried out explicitly in \cite{Bhattacharyya:2026qef}, which has subsequently been accepted for publication in this journal. That work demonstrates that the present formalism extends naturally to non-unitary Krylov dynamics, realizing the open-system extension proposed here and providing a concrete application of the framework beyond closed-system Krylov dynamics.

\begin{acknowledgments}
JM and HJRVZ
are supported in part by the ``Quantum Technologies for Sustainable Development" grant
from the National Institute for Theoretical and Computational Sciences of South Africa
(NITHECS).

\end{acknowledgments}

\bibliographystyle{JHEP}
\bibliography{biblio.bib}

\appendix

\section{Motivation for compact generating functionals}

The expression in the main text provide short and elegant expressions for the generating functions of K-complexity.  This is surprising given that, in general, only algorithmic expression for the Krylov basis and Lanczos coefficients are known.  In this appendix we show that, for a particular family of target states (for an arbitrary Hamiltonian), the generating function does indeed take a simple form and is specified purely by the Lanczos coefficients.  \\ \\
For a generic Hamiltonian (or Liouvillian), the Lanczos algorithm decomposes it as 
\begin{equation}
H = L_{+} + L_{0} + L_{-}
\end{equation}
Written in the Krylov basis, these are the usual
\begin{eqnarray}
L_{+} & = & \sum_{n} b_n |K_{n+1}\rangle \langle K_n |    \nonumber \\
L_0 & = & \sum_{n} a_n |K_{n}\rangle \langle K_n |    \nonumber \\
L_{-} & = & L_{+}^\dag    \nonumber
\end{eqnarray}
Finding the Lanczos coefficients and Krylov basis vectors is an algorithmic problem that depend sensitively on the specific Hamiltonian and reference state.  Regardless of the specifics, however, there is in every problem a class of states for which the Krylov complexity (and all higher moments) can be written in a compact way.  Consider the (unnormalized) states 
\begin{equation}
|z) = e^{z L_{+}} |K_0\rangle    \label{zStates}
\end{equation}
These are reminiscent of generalized coherent states\footnote{Strictly these are only generalized coherent states if the operators $\left\{ L_{+}, L_{-}, \left[ L_{-}, L_{+} \right]   \right\}$ close under commutation relations i.e. if we are considering a rank $1$ algebra.  } and contain all the Krylov basis vectors in a series expansion of $z$.  The overlap of these states 
\begin{eqnarray}
(\bar{z}| z) & = & \sum_{n=0}^\infty \frac{(\bar{z} z)^n}{(n!)^2} \langle K_0| L_{-}^n L_{+}^n |K_0\rangle   \nonumber \\
& = &  \sum_{n=0}^\infty \frac{(\bar{z} z)^n}{(n!)^2} \prod_{m=0}^n (b_{m})^2     \label{zOverlap}
\end{eqnarray}
contains the $b_n$ Lanczos coeffcients as a series expansion in $z$.  Furthermore, the overlap serves as a generating function for all its Krylov moments.  As an example, consider
\begin{eqnarray}
( z (\partial_z) )^m  (\bar{z}| z) & = & \sum_{n=0}^\infty n^m \frac{(\bar{z} z)^n}{(n!)^2} \langle K_0| L_{-}^n L_{+}^n |K_0\rangle     \nonumber \\
& = & \sum_{n=0}^\infty n^m \frac{(\bar{z} z)^n}{(n!)^2} \frac{\langle K_0| L_{-}^n L_{+}^n |K_0\rangle  \langle K_0| L_{-}^n L_{+}^n |K_0\rangle  }{\langle K_0| L_{-}^n L_{+}^n |K_0\rangle  }     \nonumber \\
& = & \sum_{n=0}^\infty n^m \frac{\langle K_0| e^{\bar{z} L_{-}} L_{+}^n |K_0\rangle  \langle K_0| L_{-}^n e^{z L_{+}} |K_0\rangle  }{\langle K_0| L_{-}^n L_{+}^n |K_0\rangle  }       \nonumber \\
& = & \sum_{n=0}^\infty n^m ( \bar{z} |K_n\rangle \langle K_n | z )
\end{eqnarray}
This illustrates that the states (\ref{zStates}) have simple expressions for their Krylov moments.  This is true for any Hamiltonian.  The computation of these state overlap requires detailed knowledge of the Lanczos coefficients, however.  Though the expressions are elegant the computation may thus still be difficult.  The repackaging of the continuum limit expression in terms of the Schwinger-Keldysh path integral can be understood similarly.

\section{Krylov fluctuations from the Schwinger–Keldysh loop expansion}

\noindent
In this appendix we provide a detailed derivation of the statement made in the main text that the Schwinger–Keldysh loop expansion predicts that fluctuations of Krylov complexity are parametrically enhanced when the semiclassical Krylov flow is only weakly hyperbolic over a long crossover scale. This mechanism underlies the growth of the Krylov susceptibility $\chi_K$ near integrability–chaos crossovers. Our goal is to show explicitly how this enhancement arises from the quadratic (one-loop) structure of the Schwinger–Keldysh action and why it depends on the crossover scale $n_\ast$ (or equivalently the associated escape-time $t_\ast$).\\

\noindent
We start from the SK representation of the full counting statistics
\begin{eqnarray}
    Z(\chi,t)
    =
    \int \mathcal Dn_\pm\,\mathcal Dp_\pm\;
    \exp\Bigl(
    iS[n_+,p_+]
    -
    iS[n_-,p_-]
    \Bigr),
\end{eqnarray}
with single-branch action
\begin{eqnarray}
    S[n,p]
    =
    \int_0^t dt'\,
    \big(
    p\,\dot n
    -
    H_{\rm eff}(n,p)
    \big)\,,
\end{eqnarray}
where $H_{\rm eff}(n,p)=2\,b(n)\cos p$.
Again, introducing the classical and quantum fields
\begin{eqnarray}
    n_c=\tfrac{1}{2}(n_++n_-),\quad n_q=n_+-n_-,
    \quad
    p_c=\tfrac{1}{2}(p_++p_-),
    \quad p_q=p_+-p_-\,,
\end{eqnarray}
and expanding to linear order in $(n_q,p_q)$, the Schwinger-Keldysh action takes the Martin–Siggia–Rose–Janssen–De Dominicis (MSRJD) form
\begin{eqnarray}
    S_{\rm SK}^{(1)}
    =
    \int_0^t dt\,
    \Big[
    p_q\big(\dot n_c-\partial_p H_{\rm eff}\big)
    +
    n_q\big(-\dot p_c-\partial_n H_{\rm eff}\big)
    \Big]\,.
\end{eqnarray}
Integrating over the quantum fields $(n_q,p_q)$ enforces the classical equations of motion,
\begin{eqnarray}
    \dot n_c = \,\partial_p H_{\rm eff} = -2\,b(n_c)\sin p_c,
    \qquad
    \dot p_c = -\partial_n H_{\rm eff} = -2\,b'(n_c)\cos p_c\,,
\end{eqnarray}
which, as we have seen in the main body, define the typical Krylov growth trajectory, which controls the mean complexity $K(t)=\langle n(t)\rangle$.

\subsection{Quadratic expansion and fluctuation dynamics}

To study fluctuations, we expand around the saddle trajectory,
\begin{eqnarray}
    n_c(t)=n_\ast(t)+\delta n(t),
    \qquad
    p_c(t)=p_\ast(t)+\delta p(t)\,.
\end{eqnarray}
Keeping terms quadratic in fluctuations and quantum fields, the Schwinger-Keldysh action takes the form
\begin{eqnarray}
    S_{\rm SK}^{(2)}
    =
    \int_0^t dt\,
    \Big(
    \eta_q^{\sf T}
    (\partial_t - A(t))\,\eta_c
    +
    \frac{i}{2}\,
    \eta_q^{\sf T} D(t)\,\eta_q
    \Big),
    \label{quadratic}
\end{eqnarray}
where $\eta_c=(\delta n,\delta p)^{\sf T}$, $\eta_q=(n_q,p_q)^{\sf T}$,
A(t) is the stability matrix of the classical flow, $D(t)$ is the Keldysh noise kernel, encoding quantum fluctuations. Its detailed form is model dependent, but it is smooth and nonnegative, and 
\begin{eqnarray}
    A(t)=
\begin{pmatrix}
    \partial_n\dot n & \partial_p\dot n\\
    \partial_n\dot p & \partial_p\dot p
    \end{pmatrix}_{(n_\ast(t),p_\ast(t))},
\end{eqnarray}
with $\dot n,\dot p$ given by the equations above. The sign structure in $(\partial_t-A)$ follows directly from the linearization of the equations of motion. The quadratic action \eqref{quadratic} is equivalent to a linear stochastic equation for the classical fluctuations,
\begin{eqnarray}
    \partial_t \eta_c(t)
    =
    A(t)\,\eta_c(t)
    +
    \xi(t)\,,
\end{eqnarray}
where $\langle \xi(t)\xi(t')^{\sf T}\rangle
= D(t)\,\delta(t-t')$. Now let $G(t,s)$ be the retarded Green’s function of $\partial_t-A(t)$ so that,
\begin{eqnarray}
    (\partial_t-A(t))\,G(t,s)=\delta(t-s)\,\mathbb I\,,
\end{eqnarray}
with $G(t,s)=0$ for $t<s$. The solution of this equation is given by
\begin{eqnarray}
     \eta_c(t)
     =
     \int_0^t ds\, G(t,s)\,\xi(s),
\end{eqnarray}
with associated covariance matrix
\begin{eqnarray}
    \mathrm{Cov}(t)
    =
    \int_0^t ds\;
    G(t,s)\,D(s)\,G(t,s)^{\sf T}\,.
    \label{cov-mat}
\end{eqnarray}
The second cumulant of the Krylov position is the $(n,n)$ component,
\begin{eqnarray}
    \kappa_2(t)
    =
    \mathrm{Var}[n(t)]
    =
    \big[\mathrm{Cov}(t)\big]_{nn}\,.
\end{eqnarray}
This expression makes precise the statement that fluctuations are controlled by the inverse stability operator $(\partial_t-A)^{-1}$.

\subsection{Weak hyperbolicity and parametric enhancement}
\noindent
Along the growth direction of the flow, the relevant instantaneous instability exponent is determined by the positive eigenvalue of the stability matrix $A(t)$. Parametrically,
$\lambda(t)\sim 2\,b'(n_\ast(t))$. In the fully chaotic regime $b'(n)\to\alpha>0$, the instability exponent approaches a constant and the flow is uniformly hyperbolic. In contrast, for crossover forms of $b(n)$ with a large scale $n_\ast$, $b'(n)\ll \alpha$ for
$n\lesssim n_\ast$ and the system spends a long time in a weakly hyperbolic regime. The time required for the typical trajectory to reach the linear regime scales as
\begin{eqnarray}
    t_\ast
    \sim
    \frac{1}{2\alpha}\log n_\ast\, .
\end{eqnarray}
Since the propagator $G(t,s)$ decays only weakly for $0<s<t_\ast$, the integral in \eqref{cov-mat} accumulates contributions over the entire crossover interval. Parametrically then,
\begin{eqnarray}
    \kappa_2(t)
    \;\propto\;
    \int_0^{t_\ast} ds\,\|G(t,s)\|^2
    \;\sim\;
    t_\ast , 
\end{eqnarray}
up to nonuniversal prefactors set by the noise kernel $D(t)$. In other words, the variance, and hence the Krylov susceptibility, grows with the duration of weak hyperbolicity, not merely with the asymptotic growth rate. The Schwinger-Keldysh loop expansion makes it explicit that Krylov fluctuations are governed by the stability of the semiclassical flow, and that near integrability–chaos crossovers, where hyperbolicity is weak over long scales, these fluctuations are parametrically enhanced even when the mean complexity varies smoothly.

\end{document}